\documentclass[onecolumn,10pt,aps,prd,preprintnumbers,superscriptaddress,showkeys,
nofootinbib,amsmath,amssymb,floatfix,longbibliography]{revtex4-2}

\usepackage{bm}
\usepackage{booktabs}
\usepackage{graphicx}
\makeatletter
\def\label#1{\@bsphack
     \begingroup
     \UseHookWithArguments{label}{1}{#1}%
     \protected@write\@auxout{}%
         {\string\newlabel{#1}{{\@currentlabel}{\thepage}%
         {\@currentlabelname}{\@currentHref}{\@kernel@reserved@label@data}}}%
  \endgroup
  \@esphack}
\def\@bibdataout@aps{%
 \immediate\write\@bibdataout{%
  @CONTROL{%
   apsrev42Control%
   \longbibliography@sw{%
    ,author="48",editor="1",pages="0",title="0",year="1"%
   }{%
    ,author="48",editor="1",pages="0",title="",year="1"%
   }%
  }%
 }%
 \if@filesw
  \immediate\write\@auxout{\string\citation{apsrev42Control}}%
 \fi
}
\makeatother
\usepackage{hyperref}
\usepackage{microtype}

\hypersetup{colorlinks=true,linkcolor=blue,citecolor=blue,urlcolor=blue}
\allowdisplaybreaks

\newcommand{\dd}{\mathrm{d}}

\newcommand{\order}{\mathcal{O}}
\newcommand{\identity}{\mathbb{I}}

\begin{document}

\title{Two-branch detector response for Dirac infall into a Schwarzschild--MOG black hole}

\author{Nikko John Leo S. Lobos}
\email{nikko\_john\_s\_lobos@dlsu.edu.ph}
\affiliation{Department of Physics, De La Salle University, 2401 Taft Avenue, Manila 1004, Philippines}

\author{Emmanuel T. Rodulfo}
\email{emmanuel.rodulfo@dlsu.edu.ph}
\affiliation{Department of Physics, De La Salle University, 2401 Taft Avenue, Manila 1004, Philippines}

\begin{abstract}
We study a localized spin-$1/2$ detector falling into a Schwarzschild--MOG black hole. The detector interacts with a neutral massless scalar field through a charge-preserving two-level transition, while its translational wave packet obeys the MOG-charged Dirac equation. Near the outer horizon, the separated radial system reduces to an inverse-square equation with index $\Theta_{\rm D}=E_{\rm H}/(2\hbar\kappa_\alpha)$. The gauge-invariant horizon energy satisfies $E_{\rm H}=m_\psi U_{\rm H}>0$ for every future-directed crossing trajectory. We quantize the scalar field in a globally normalized Boulware scattering basis and retain both radial-flux branches of a mode that is outgoing at infinity. This treatment does not identify a local outgoing ansatz with a complete mode. A finite radial gate $\chi_p(x)=(x/L_\chi)^p e^{-x/L_\chi}$ gives closed-form excitation and absorption probability densities that include the ingoing contribution and scattering-phase interference. Within the controlled near-horizon approximation, the full detailed-balance ratio factorizes into a branch-resolved outgoing ratio and a two-branch factor. Only the outgoing ratio approaches $\exp(-2\pi\nu/\kappa_\alpha)$ when the near-horizon, adiabatic, high-gap, and branch-isolation conditions all hold. The leading switching correction is controlled by $(\nu/\kappa_\alpha)/(S_\pm L_\chi)$, not by $(S_\pm L_\chi)^{-1}$ alone. In a weak-MOG expansion, the local correction separates into surface-gravity, trajectory-prefactor, and finite-gate terms. The full response also contains a contribution from global scattering. This separation shows which terms follow from the local horizon geometry and which depend on the detector protocol and scalar propagation outside the horizon region.
\end{abstract}


\maketitle

\section{Introduction}
\label{sec:introduction}

Quantum field theory on curved spacetime does not generally provide a unique particle definition. Particle content depends on the positive-frequency splitting, the quantum state, and the observer. This dependence underlies cosmological particle creation, the response of accelerated detectors, and Hawking emission \cite{Parker:1968ParticleCreation,Parker:1969QuantizedFields,Fulling:1973Nonuniqueness,Davies:1975ScalarProduction,Unruh:1976db,Hawking:1975vcx}. A localized two-level detector makes the observer dependence operational. Its transition probability is determined by the field state, its worldline, its energy gap, and the way in which the interaction is switched on and off \cite{DeWitt:1979QuantumGravity,Louko:Satz2008}.

Horizon-brightened acceleration radiation (HBAR) applies this detector model to atoms falling toward a black-hole horizon. In the original construction, an atom crosses a mode-selective cavity in the Schwarzschild exterior and can become excited while emitting a scalar quantum through the counter-rotating term of the atom--field interaction \cite{Scully:2017utk}. The logarithmic phase of an outgoing field branch along the infalling worldline produces the Planck factor. Near-horizon conformal quantum mechanics gives the same origin for this factor \cite{Camblong:2020qtd}. Master-equation and entropy analyses have also examined how repeated atomic injection can produce a thermal steady state and an entropy flux analogous to that of a black hole \cite{Azizi:2021Master,Azizi:2021Thermodynamics}.

The HBAR mechanism has been extended to general static metrics, rotating and charged backgrounds, braneworld geometries, quantum-corrected black holes, derivative couplings, and massive vector fields \cite{Sen:2022static,Azizi:2021KerrHBAR,Sen:2023kerrnewman,Das:2024BraneworldHBAR,Jana:2024ChargedHBAR,Das:2025DerivativeHBAR,Pantig:2026VectorHBAR}. In these calculations, a nondegenerate horizon fixes the leading logarithmic exponent. The absolute probability also depends on the trajectory, interaction operator, field content, switching, boundary conditions, and global propagation. This distinction matters for finite measurements because switching transients can dominate the response of a freely falling detector near a horizon \cite{ShallueCarroll:2025}.

Two properties of the scalar modes are relevant here. A local solution proportional to $e^{-i\nu(t-r_*)}$ represents only one asymptotic branch. A globally normalized scattering mode that is outgoing at infinity generally contains both signs of radial Klein--Gordon flux near the horizon. A Wronskian identity relates their coefficients, which contain greybody information \cite{Candelas:1980,HodgkinsonLoukoOttewill:2014,Law:2022Scattering}. A frequency filter also does not select a radial direction. When a detector couples locally to the complete mode, both branches and their interference contribute to the transition probability. A one-branch HBAR formula therefore describes either a direction-resolved observable or an approximation with a stated error.

Modified gravity also changes the detector trajectory. Scalar--tensor--vector gravity (STVG), also called modified gravity (MOG), contains a vector field that couples universally to massive matter \cite{Moffat:2005si,Moffat:2014aja,Moffat:2021MOG}. In the constant-scalar, massless-vector black-hole sector, the geometry is Reissner--Nordstr\"om-like, but the vector charge is gravitational rather than electromagnetic. A massive test body carries a mass-proportional MOG charge and follows a Lorentz-force-type trajectory. Replacing the Schwarzschild surface gravity with its MOG value while retaining a Schwarzschild geodesic omits the vector-force contribution from the same theory.

We study these effects with a localized Dirac detector. Its center-of-mass packet solves the MOG-charged Dirac equation, and an independent two-level degree of freedom supplies the detector gap. In this effective model, the translational mass and MOG charge are fixed. We neglect recoil and state-dependent changes of the worldline when $\hbar\omega/m_\psi\ll1$. The emitted quantum is a neutral scalar, and the interaction is proportional to the identity in translational spin space. This model does not describe coupling to a quantized fermion field and does not predict a Fermi--Dirac spectrum, spin flip, or helicity asymmetry \cite{Louko:2016FermionDetector}.

We derive the separated radial Dirac equations and their controlled near-horizon reduction. For the same WKB packet, the horizon Dirac energy and the horizon trajectory function obey
\begin{equation}
E_{\rm H}=E-q_\psi\Phi_{\rm H}=m_\psi U_{\rm H}.
\label{eq:intro_energy_identity}
\end{equation}
Here $E$ is the canonical Killing energy, $q_\psi$ is the detector's MOG charge, and $\Phi_{\rm H}$ is the horizon potential difference. The formal Dirac threshold $E_{\rm H}=0$ is therefore also the trajectory threshold $U_{\rm H}=0$. This threshold does not describe a future-directed horizon-crossing HBAR trajectory.

We then use a globally normalized scalar mode that is outgoing at infinity and retain both of its near-horizon branches. A finite radial gate makes the endpoint integrals convergent and allows the response to be evaluated in closed form. If $q_{\rm exc}$ and $q_{\rm abs}$ denote the ingoing-to-outgoing amplitude ratios for excitation and absorption, respectively, the exact probability-density ratio takes the form
\begin{equation}
\frac{\dd P_{\rm exc}/\dd\nu}{\dd P_{\rm abs}/\dd\nu}
=\mathcal{R}_{\rm out}(\nu)
\frac{|1+q_{\rm exc}(\nu)|^2}{|1+q_{\rm abs}(\nu)|^2}.
\label{eq:intro_full_factorization}
\end{equation}
The factor $\mathcal{R}_{\rm out}$ is the branch-resolved outgoing response. The second factor depends on the global reflection coefficient and the regular ingoing trajectory integral. The thermal one-branch law equals the full response of a minimally coupled detector only when this second factor is close to unity.

For a weak MOG deformation, the logarithmic detailed-balance function separates into local and global terms. The local part contains surface-gravity, trajectory-prefactor, and finite-gate contributions. The global term is the derivative of the two-branch factor in Eq.~\eqref{eq:intro_full_factorization} and cannot be inferred from the near-horizon geometry alone.

Equations~\eqref{eq:background_fields}--\eqref{eq:test_charge} apply the established STVG black-hole and matter prescriptions. Equations~\eqref{eq:covariant_dirac_equation}, \eqref{eq:scalar_radial}, and \eqref{eq:interaction} use standard curved-spacetime Dirac, scalar-mode, and monopole-detector formulations. We derive the charged separated system, its threshold behavior, the MOG-forced near-horizon expansions, the finite-gate transforms, the coherent two-branch response, and the weak-MOG decomposition. We compare the branch-isolated Schwarzschild limit with the original HBAR calculation and its conformal near-horizon formulation \cite{Scully:2017utk,Camblong:2020qtd}.

Section~\ref{sec:background} defines the Schwarzschild--MOG background and the effective detector. Section~\ref{sec:dirac} derives the radial Dirac system, its inverse-square limit, the threshold solution, and the MOG-forced trajectory. Section~\ref{sec:scalar} defines the global scalar channel and finite radial gate. Section~\ref{sec:response} gives the two-branch probabilities and the controlled detailed-balance limit. Section~\ref{sec:limits} treats the Schwarzschild and weak-MOG limits, observable ratios, and validity conditions. Section~\ref{sec:discussion} compares the results with earlier HBAR work and summarizes the limitations.

\section{Schwarzschild--MOG background and effective detector}
\label{sec:background}

We use the signature $(-,+,+,+)$, set $c=1$, and retain $\hbar$ and Boltzmann's constant $k_{\rm B}$. The constant-scalar, massless-vector Schwarzschild--MOG exterior is specified by
\begin{subequations}
\label{eq:background_fields}
\begin{align}
\dd s^2&=-f_\alpha(r)\dd t^2+\frac{\dd r^2}{f_\alpha(r)}+r^2\dd\Omega^2,
\label{eq:background_metric}\\
f_\alpha(r)&=1-\frac{2G_{\rm N}(1+\alpha)M}{r}
+\frac{G_{\rm N}^2\alpha(1+\alpha)M^2}{r^2},
\label{eq:background_lapse}\\
\phi_\mu\dd x^\mu&=-\frac{Q_{\rm MOG}}{r}\dd t,
\qquad Q_{\rm MOG}=\sqrt{\alpha G_{\rm N}}\,M.
\label{eq:background_vector}
\end{align}
\end{subequations}
The mass parameter is $M>0$, Newton's constant is $G_{\rm N}$, and the dimensionless deformation parameter satisfies $\alpha\geq0$. The gauge in Eq.~\eqref{eq:background_vector} is fixed by $\phi_t(\infty)=0$. We use this black-hole sector without dynamically varying the scalar fields \cite{Moffat:2014aja,Mureika:2015sfa,Moffat:2021MOG}.

The lapse factorizes as $f_\alpha=(r-r_+)(r-r_-)/r^2$, with
\begin{subequations}
\label{eq:horizon_data}
\begin{align}
r_\pm&=G_{\rm N}M\left(1+\alpha\pm\sqrt{1+\alpha}\right),
\label{eq:horizon_radii}\\
\kappa_\alpha&=\frac{f_\alpha'(r_+)}{2}
=\frac{r_+-r_-}{2r_+^2},
\qquad
T_{\rm H}^{\rm MOG}=\frac{\hbar\kappa_\alpha}{2\pi k_{\rm B}}.
\label{eq:surface_gravity_temperature}
\end{align}
\end{subequations}
Equation~\eqref{eq:horizon_radii} gives $r_+-r_-=2G_{\rm N}M\sqrt{1+\alpha}>0$ for every finite $\alpha\geq0$, so the outer horizon is nondegenerate throughout the range considered. Equation~\eqref{eq:surface_gravity_temperature} fixes the Killing normalization using the asymptotically unit time coordinate $t$. These are the standard Schwarzschild--MOG horizon expressions \cite{Moffat:2014aja,Mureika:2015sfa}.

The horizon potential difference is
\begin{equation}
\Phi_{\rm H}=\phi_t(\infty)-\phi_t(r_+)=\frac{Q_{\rm MOG}}{r_+}.
\label{eq:horizon_potential}
\end{equation}
Only this potential difference and the corresponding canonical energy enter gauge-invariant results.

The translational detector is a localized spinor packet of reference mass $m_\psi$ and standard MOG charge
\begin{equation}
q_\psi=\sqrt{\alpha G_{\rm N}}\,m_\psi,
\qquad
\zeta_\psi\equiv\frac{q_\psi}{m_\psi}=\sqrt{\alpha G_{\rm N}}.
\label{eq:test_charge}
\end{equation}
This proportionality is part of the STVG matter prescription. The vector force therefore cannot be removed independently while the MOG background is held fixed \cite{Moffat:2014aja,Moffat:2021MOG}.

With the Clifford convention $\{\gamma^a,\gamma^b\}=-2\eta^{ab}\identity_4$ for $\eta_{ab}=\mathrm{diag}(-1,1,1,1)$, the spinor obeys
\begin{equation}
\left(i\hbar\gamma^a e_a{}^\mu\mathcal{D}_\mu-m_\psi\right)\psi=0,
\qquad
\mathcal{D}_\mu=\partial_\mu+\Gamma_\mu-\frac{i}{\hbar}q_\psi\phi_\mu.
\label{eq:covariant_dirac_equation}
\end{equation}
Here $e_a{}^\mu$ is an inverse tetrad and $\Gamma_\mu=\omega_{\mu ab}[\gamma^a,\gamma^b]/8$ is its spin connection. This convention fixes the sign of the covariant energy used in the radial and WKB equations and follows the standard tetrad formulation of the Dirac equation in curved spacetime \cite{Alcubierre:2025dirac}.

The detector's internal Hilbert space is spanned by a ground state $|b\rangle$ and an excited state $|a\rangle$ with proper energy gap $\hbar\omega>0$. The internal monopole operator is
\begin{equation}
\widehat\mu(\tau)=\sigma^-e^{-i\omega\tau}+\sigma^+e^{i\omega\tau},
\qquad
\sigma^+=|a\rangle\langle b|,
\quad
\sigma^-=|b\rangle\langle a|.
\label{eq:monopole_operator}
\end{equation}
Both internal states belong to the same translational packet and carry the same conserved MOG charge. We do not treat $\hbar\omega$ as a state-dependent change in $m_\psi$ within the force law. We neglect this backreaction and recoil under the condition
\begin{equation}
\epsilon_{\rm rec}\equiv\frac{\hbar\omega}{m_\psi}\ll1.
\label{eq:recoil_condition}
\end{equation}
This condition allows one reference worldline to describe both internal levels. A microscopic composite detector beyond this order would require a specified internal stress-energy tensor and MOG current.

The emitted field is a real, massless scalar $\widehat\Phi$ with no MOG charge. Its action is $S_\Phi=-\frac12\int\sqrt{-g}\,g^{\mu\nu}\nabla_\mu\Phi\nabla_\nu\Phi\,\dd^4x$, which gives $\Box\Phi=0$. We define the detector--field interaction after introducing the globally normalized scalar channel and finite radial gate. The interaction is proportional to the identity in translational spin space, so the Dirac spin labels do not change during the internal transition.

\section{Dirac propagation and the MOG-forced trajectory}
\label{sec:dirac}

For the diagonal coframe $e^0=\sqrt{f_\alpha}\,\dd t$, $e^1=\dd r/\sqrt{f_\alpha}$, $e^2=r\dd\theta$, and $e^3=r\sin\theta\,\dd\varphi$, Eq.~\eqref{eq:covariant_dirac_equation} becomes
\begin{equation}
\begin{aligned}
0={}&\Bigg[
\frac{i\hbar\gamma^0}{\sqrt{f_\alpha}}
\left(\partial_t-\frac{i}{\hbar}q_\psi\phi_t\right)
+i\hbar\gamma^1\sqrt{f_\alpha}
\left(\partial_r+\frac1r+\frac{f_\alpha'}{4f_\alpha}\right)\\
&+\frac{i\hbar\gamma^2}{r}\left(\partial_\theta+\frac12\cot\theta\right)
+\frac{i\hbar\gamma^3}{r\sin\theta}\partial_\varphi-m_\psi\Bigg]\psi.
\end{aligned}
\label{eq:explicit_dirac}
\end{equation}
The terms $1/r$ and $f_\alpha'/(4f_\alpha)$ in Eq.~\eqref{eq:explicit_dirac} come from the radial spin connection. They must be retained before taking the horizon limit because removing them requires a specific spinor rescaling \cite{Alcubierre:2025dirac}.

Introduce the tortoise coordinate $\dd r_*/\dd r=f_\alpha^{-1}$ and write $\psi=f_\alpha^{-1/4}r^{-1}\widetilde\psi$. The stationary spinor is separated as
\begin{equation}
\widetilde\psi_{E\varkappa m}
=e^{-iEt/\hbar}
\begin{pmatrix}
F_\varkappa(r)\Omega_{\varkappa m}\\[2pt]
iG_\varkappa(r)\Omega_{-\varkappa m}
\end{pmatrix},
\qquad
\varkappa=\pm\left(j+\frac12\right).
\label{eq:spinor_separation}
\end{equation}
Here $\Omega_{\varkappa m}$ are spinor spherical harmonics, $j$ is the total angular momentum, and $m$ is its azimuthal projection. A constant phase convention for $G_\varkappa$ gives the exact radial system
\begin{subequations}
\label{eq:radial_dirac_system}
\begin{align}
\left(\frac{\dd}{\dd r_*}+\frac{\varkappa\sqrt{f_\alpha}}{r}\right)F_\varkappa
&=-\left(\frac{\mathcal{E}}{\hbar}+\frac{m_\psi\sqrt{f_\alpha}}{\hbar}\right)G_\varkappa,
\label{eq:radial_F}\\
\left(\frac{\dd}{\dd r_*}-\frac{\varkappa\sqrt{f_\alpha}}{r}\right)G_\varkappa
&=\left(\frac{\mathcal{E}}{\hbar}-\frac{m_\psi\sqrt{f_\alpha}}{\hbar}\right)F_\varkappa,
\label{eq:radial_G}\\
\mathcal{E}(r)&=E+q_\psi\phi_t(r)=E-\frac{q_\psi Q_{\rm MOG}}{r}.
\label{eq:radial_energy}
\end{align}
\end{subequations}
Equations~\eqref{eq:radial_F} and \eqref{eq:radial_G} retain the mass, angular, vector-potential, and metric terms. Up to the stated component-phase convention, their uncharged Schwarzschild structure agrees with standard massive-Dirac radial systems \cite{ChoLin:2005Dirac}. The replacement $E\mapsto\mathcal{E}(r)$ and the MOG lapse give the charged STVG specialization used here. The horizon frequency is the gauge-covariant quantity $E_{\rm H}=\mathcal{E}(r_+)=E-q_\psi\Phi_{\rm H}$.

Define $D_*=\dd/\dd r_*$, $W_\varkappa=\varkappa\sqrt{f_\alpha}/r$, $\epsilon=\mathcal{E}/\hbar$, and $\mu=m_\psi\sqrt{f_\alpha}/\hbar$. Eliminating one component gives
\begin{subequations}
\label{eq:decoupled_dirac_system}
\begin{align}
\left[(D_*-W_\varkappa)\frac{1}{\epsilon+\mu}(D_*+W_\varkappa)+\epsilon-\mu\right]F_\varkappa&=0,
\label{eq:decoupled_dirac_F}\\
\left[(D_*+W_\varkappa)\frac{1}{\epsilon-\mu}(D_*-W_\varkappa)+\epsilon+\mu\right]G_\varkappa&=0.
\label{eq:decoupled_dirac_G}
\end{align}
\end{subequations}
These exact decoupled forms hold wherever the displayed denominators do not vanish. Radial derivatives of $W_\varkappa$, $\epsilon$, and $\mu$ remain inside the ordered operators and cannot be discarded before specifying an approximation.

Let $x=r-r_+$ in the exterior. The background expansions are
\begin{subequations}
\label{eq:near_horizon_background_expansions}
\begin{align}
f_\alpha&=2\kappa_\alpha x+\frac12f''_+x^2+\order(x^3),
\label{eq:lapse_horizon_expansion}\\
\mathcal{E}&=E_{\rm H}+\frac{q_\psi Q_{\rm MOG}}{r_+^2}x+\order(x^2).
\label{eq:radial_energy_horizon_expansion}
\end{align}
\end{subequations}
where $f''_+=f_\alpha''(r_+)$. The angular and mass functions scale as $W_\varkappa=\varkappa\sqrt{2\kappa_\alpha x}/r_++\order(x^{3/2})$ and $\mu=m_\psi\sqrt{2\kappa_\alpha x}/\hbar+\order(x^{3/2})$. For $E_{\rm H}\neq0$, a sufficient control condition is
\begin{equation}
\delta_{\rm D}(x)
\equiv\frac{\sqrt{f_\alpha(r)}}{|E_{\rm H}|}
\left(m_\psi+\frac{\hbar|\varkappa|}{r_+}\right)\ll1,
\qquad
\frac{|q_\psi Q_{\rm MOG}|x}{r_+^2|E_{\rm H}|}\ll1.
\label{eq:dirac_control}
\end{equation}
These inequalities ensure that the constant horizon energy dominates the redshifted mass, angular-momentum, and potential-gradient terms.

Under these conditions, the combinations $Z_{\rm out}=F_\varkappa+iG_\varkappa$ and $Z_{\rm in}=F_\varkappa-iG_\varkappa$ obey $\dd Z_{\rm out,in}/\dd r_*\simeq\pm iE_{\rm H}Z_{\rm out,in}/\hbar$. Since
\begin{equation}
r_*=\frac{1}{2\kappa_\alpha}\ln\left(\frac{x}{r_+}\right)
-\frac{f''_+}{8\kappa_\alpha^2}x+\order(x^2),
\label{eq:tortoise_horizon_expansion}
\end{equation}
the two leading branches are
\begin{subequations}
\label{eq:dirac_horizon_branches}
\begin{align}
Z_{\rm out}&=C_{\rm out}\left(\frac{x}{r_+}\right)^{i\Theta_{\rm D}}
\left[1+\order(\delta_{\rm D})+\order(x/r_+)\right],
\label{eq:dirac_out_branch}\\
Z_{\rm in}&=C_{\rm in}\left(\frac{x}{r_+}\right)^{-i\Theta_{\rm D}}
\left[1+\order(\delta_{\rm D})+\order(x/r_+)\right],
\qquad
\Theta_{\rm D}=\frac{E_{\rm H}}{2\hbar\kappa_\alpha}.
\label{eq:dirac_in_branch}
\end{align}
\end{subequations}
The index $\Theta_{\rm D}$ is the ratio of the horizon-frame packet energy to the surface-gravity scale.

Applying a second derivative and defining $U_{\rm D}=\sqrt{x}\,Z_{\rm out,in}$ yields the inverse-square equation
\begin{equation}
\frac{\dd^2U_{\rm D}}{\dd x^2}
+\frac{\lambda_{\rm D}}{x^2}U_{\rm D}\simeq0,
\qquad
\lambda_{\rm D}=\frac14+\Theta_{\rm D}^2.
\label{eq:dirac_cqm}
\end{equation}
Equation~\eqref{eq:dirac_cqm} is the near-horizon conformal form of the translational spinor equation. It differs from the scalar HBAR response equation. The scalar inverse-square reduction used in HBAR is discussed in Refs.~\cite{Camblong:2020qtd,Sen:2022static}, whereas Eq.~\eqref{eq:dirac_cqm} is its spinor counterpart. The omitted terms begin at relative order $\sqrt{x}$ and are controlled by Eq.~\eqref{eq:dirac_control}.

The threshold $E_{\rm H}=0$ requires a separate expansion because the terms proportional to $\sqrt{f_\alpha}$ then dominate the energy term. Write $\epsilon=\mathcal{E}_1x+\order(x^2)$, $W_\varkappa=\mathcal{W}_\varkappa\sqrt{x}+\order(x^{3/2})$, and $\mu=\mathcal{M}_\psi\sqrt{x}+\order(x^{3/2})$, with
\begin{subequations}
\label{eq:threshold_scales}
\begin{align}
\mathcal{E}_1&=\frac{q_\psi Q_{\rm MOG}}{\hbar r_+^2},
\label{eq:threshold_energy_scale}\\
\mathcal{W}_\varkappa&=\frac{\varkappa\sqrt{2\kappa_\alpha}}{r_+},
\label{eq:threshold_angular_scale}\\
\mathcal{M}_\psi&=\frac{m_\psi\sqrt{2\kappa_\alpha}}{\hbar}.
\label{eq:threshold_mass_scale}
\end{align}
\end{subequations}
A regular Frobenius series, $F=F_0+F_1\sqrt{x}+F_2x+\cdots$ and $G=G_0+G_1\sqrt{x}+G_2x+\cdots$, gives
\begin{subequations}
\label{eq:threshold_series_coefficients}
\begin{align}
F_1&=-\frac{\mathcal{W}_\varkappa F_0+\mathcal{M}_\psi G_0}{\kappa_\alpha},
\label{eq:threshold_F1}\\
G_1&=\frac{\mathcal{W}_\varkappa G_0-\mathcal{M}_\psi F_0}{\kappa_\alpha},
\label{eq:threshold_G1}\\
F_2&=\frac{\mathcal{W}_\varkappa^2+\mathcal{M}_\psi^2}{2\kappa_\alpha^2}F_0
-\frac{\mathcal{E}_1}{2\kappa_\alpha}G_0,
\label{eq:threshold_F2}\\
G_2&=\frac{\mathcal{W}_\varkappa^2+\mathcal{M}_\psi^2}{2\kappa_\alpha^2}G_0
+\frac{\mathcal{E}_1}{2\kappa_\alpha}F_0.
\label{eq:threshold_G2}
\end{align}
\end{subequations}
The mass and angular terms determine the first $\sqrt{x}$ correction. The vector-potential gradient first enters at order $x$. Setting $\Theta_{\rm D}=0$ directly in Eq.~\eqref{eq:dirac_cqm} would miss this change in the dominant balance.

The WKB limit connects the spinor phase to the packet trajectory and gives
\begin{equation}
g^{\mu\nu}(\partial_\mu S-q_\psi\phi_\mu)
(\partial_\nu S-q_\psi\phi_\nu)+m_\psi^2=0.
\label{eq:wkb_hamilton_jacobi}
\end{equation}
Consequently, the center of a narrow positive-energy packet obeys $u^\nu\nabla_\nu u^\mu=\zeta_\psi B^\mu{}_{\nu}u^\nu$, where $B_{\mu\nu}=\nabla_\mu\phi_\nu-\nabla_\nu\phi_\mu$. For radial motion, let $\varepsilon=E/m_\psi$ and define
\begin{equation}
U(r)=\varepsilon+\zeta_\psi\phi_t(r)
=\varepsilon-\frac{\alpha G_{\rm N}M}{r}.
\label{eq:trajectory_U}
\end{equation}
The future-directed infalling branch satisfies
\begin{subequations}
\label{eq:trajectory_equations}
\begin{align}
\frac{\dd r}{\dd\tau}&=-\sqrt{U^2(r)-f_\alpha(r)},
\label{eq:trajectory_r}\\
\frac{\dd t}{\dd\tau}&=\frac{U(r)}{f_\alpha(r)},
\qquad
U_{\rm H}\equiv U(r_+)>0.
\label{eq:trajectory_t}
\end{align}
\end{subequations}
Equations~\eqref{eq:trajectory_r} and \eqref{eq:trajectory_t} are the radial first integrals of the STVG Lorentz-force-type equation for $q_\psi/m_\psi=\sqrt{\alpha G_{\rm N}}$ \cite{Moffat:2014aja,Moffat:2021MOG}. They include the repulsive MOG vector force. Below, ``infall'' refers to this forced trajectory rather than a metric geodesic.

Using Eq.~\eqref{eq:test_charge}, the horizon quantities obey the exact identity
\begin{equation}
E_{\rm H}=E-q_\psi\Phi_{\rm H}
=m_\psi\left(\varepsilon-\zeta_\psi\Phi_{\rm H}\right)
=m_\psi U_{\rm H}.
\label{eq:horizon_identity}
\end{equation}
Thus every physical crossing trajectory has $E_{\rm H}>0$. The stationary threshold series lies at $U_{\rm H}=0$, where the timelike exterior trajectory and the proper-time expansion used for HBAR cease to exist.

The response calculation requires the trajectory phase through quadratic order. Define $U_1=U'(r_+)=\alpha G_{\rm N}M/r_+^2$, $f''_+=f_\alpha''(r_+)$, and $f'''_+=f_\alpha'''(r_+)$. Expanding Eq.~\eqref{eq:trajectory_equations} and $r_*$ gives
\begin{subequations}
\label{eq:null_trajectory_expansion}
\begin{align}
\tau(x)&=\tau_{\rm H}-\frac{x}{U_{\rm H}}+T_2x^2+\order(x^3),
&T_2&=\frac{U_1}{2U_{\rm H}^2}-\frac{\kappa_\alpha}{2U_{\rm H}^3},
\label{eq:tau_expansion}\\
u(x)&=u_0-\frac{1}{\kappa_\alpha}\ln\left(\frac{x}{r_+}\right)-C_ux+D_ux^2+\order(x^3),
\label{eq:u_expansion}\\
v(x)&=v_0-\frac{x}{2U_{\rm H}^2}+D_vx^2+\order(x^3),
\label{eq:v_expansion}
\end{align}
\end{subequations}
where $u=t-r_*$ and $v=t+r_*$ are retarded and advanced times. Their coefficients are
\begin{subequations}
\label{eq:null_trajectory_coefficients}
\begin{align}
C_u&=\frac{1}{2U_{\rm H}^2}-\frac{f''_+}{4\kappa_\alpha^2},
\label{eq:coefficient_Cu}\\
D_u&=-\frac{(f''_+)^2}{32\kappa_\alpha^3}
+\frac{f'''_+}{24\kappa_\alpha^2}
+\frac{U_1}{2U_{\rm H}^3}-\frac{3\kappa_\alpha}{8U_{\rm H}^4},
\label{eq:coefficient_Du}\\
D_v&=\frac{U_1}{2U_{\rm H}^3}-\frac{3\kappa_\alpha}{8U_{\rm H}^4}.
\label{eq:coefficient_Dv}
\end{align}
\end{subequations}
The retarded-time expansion in Eq.~\eqref{eq:u_expansion} contains the logarithm that produces the outgoing HBAR factor, whereas the advanced-time expansion in Eq.~\eqref{eq:v_expansion} is regular at the future horizon. The quadratic coefficients provide a frequency-dependent estimate of the phase-truncation error.

\section{Global scalar channel and finite radial gate}
\label{sec:scalar}

A positive-frequency neutral scalar mode is written as
\begin{equation}
u_{\nu\ell m}^{(s)}(x)
=\mathcal{N}_{\nu\ell}^{(s,\alpha)}\frac{e^{-i\nu t}}{r}
Y_{\ell m}(\theta,\varphi)\psi_{\nu\ell}^{(s)}(r),
\qquad \nu>0,
\label{eq:scalar_mode}
\end{equation}
where $s$ labels a global scattering channel and $\mathcal{N}_{\nu\ell}^{(s,\alpha)}$ is fixed by the Klein--Gordon inner product. We retain its possible background dependence because it enters MOG-to-Schwarzschild ratios. Substitution into $\Box\Phi=0$ gives
\begin{equation}
\frac{\dd^2\psi_{\nu\ell}^{(s)}}{\dd r_*^2}
+\left[\nu^2-V_\ell(r)\right]\psi_{\nu\ell}^{(s)}=0,
\qquad
V_\ell=f_\alpha\left[\frac{\ell(\ell+1)}{r^2}+\frac{f_\alpha'}{r}\right].
\label{eq:scalar_radial}
\end{equation}
Equation~\eqref{eq:scalar_radial} is the standard one-dimensional scalar scattering equation for a static spherical black hole \cite{Candelas:1980,HodgkinsonLoukoOttewill:2014}. The potential vanishes at the outer horizon and at spatial infinity. The barrier between these limits determines the nonlocal reflection and transmission data.

For the selected $s$ wave, $V_0=4\kappa_\alpha^2x/r_++\order(x^2)$. Factoring out the exact plane-wave phases and substituting $\psi^{\rm o,i}=e^{\pm i\nu r_*}[1+c_V^{\rm o,i}x+\order(x^2)]$ into Eq.~\eqref{eq:scalar_radial} gives
\begin{equation}
c_V^{\rm o}=\frac{\kappa_\alpha}{r_+(\kappa_\alpha+i\nu)},
\qquad
c_V^{\rm i}=\frac{\kappa_\alpha}{r_+(\kappa_\alpha-i\nu)}.
\label{eq:scalar_horizon_correction}
\end{equation}
Equation~\eqref{eq:scalar_horizon_correction} gives the first local correction to the constant branch amplitudes. Its contribution to an endpoint integral is controlled by $|c_V^b|\mathfrak m_{b,\pm}^{(1)}$, where $\mathfrak m_{b,\pm}^{(1)}$ is defined in Eq.~\eqref{eq:endpoint_moments}. The pointwise ratio $|V_0|/\nu^2$ alone does not control this contribution.

We use an $s$-wave mode that is purely outgoing at future null infinity. With unit radial amplitude at infinity, its asymptotic conditions are
\begin{subequations}
\label{eq:out_mode_boundaries}
\begin{align}
\psi_{\nu0}^{\rm out}&\longrightarrow e^{i\nu r_*},
&&r_*\longrightarrow+\infty,
\label{eq:out_at_infinity}\\
\psi_{\nu0}^{\rm out}&\longrightarrow
A_{\nu\alpha}^{\rm o}e^{i\nu r_*}
+A_{\nu\alpha}^{\rm i}e^{-i\nu r_*},
&&r_*\longrightarrow-\infty.
\label{eq:out_at_horizon}
\end{align}
\end{subequations}
The superscripts ${\rm o}$ and ${\rm i}$ in Eq.~\eqref{eq:out_at_horizon} denote positive and negative radial flux at the horizon. They are not two independent final particles, but components of the same global mode in Eq.~\eqref{eq:out_at_infinity}.

For a radial solution $\psi$, the conserved flux is $\mathcal{J}_{r_*}=(\psi^*\psi'-\psi\psi'^*)/(2i)$. Flux conservation between the two ends gives
\begin{equation}
|A_{\nu\alpha}^{\rm o}|^2-|A_{\nu\alpha}^{\rm i}|^2=1.
\label{eq:wronskian_identity}
\end{equation}
Equation~\eqref{eq:wronskian_identity} is the conserved radial-flux relation for the selected scattering solution \cite{Candelas:1980,Law:2022Scattering}. A normalization choice cannot remove the second horizon branch. With $\mathcal{T}_{\nu0}^{(\alpha)}=1/A_{\nu\alpha}^{\rm o}$ and $\mathcal{R}_{\nu0}^{(\alpha)}=A_{\nu\alpha}^{\rm i}/A_{\nu\alpha}^{\rm o}$, the relation becomes $|\mathcal{T}_{\nu0}^{(\alpha)}|^2+|\mathcal{R}_{\nu0}^{(\alpha)}|^2=1$. The greybody factor is $\Gamma_{\nu0}^{(\alpha)}=|\mathcal{T}_{\nu0}^{(\alpha)}|^2$.

The globally normalized modes satisfy
\begin{equation}
(u_{\nu\ell m}^{(s)},u_{\nu'\ell'm'}^{(s')})_{\rm KG}
=\delta_{ss'}\delta(\nu-\nu')\delta_{\ell\ell'}\delta_{mm'},
\label{eq:kg_mode_normalization}
\end{equation}
and their operators obey the corresponding continuum commutator. The exterior vacuum $|0_{\rm B}\rangle$ contains no positive Killing-frequency quanta. It is the Boulware state associated with the static exterior basis \cite{Boulware:1975QFT,Candelas:1980}. Although this state is singular on the horizon, the interaction is finite, switched, and restricted to the exterior, as in the HBAR state prescription \cite{Scully:2017utk,Camblong:2020qtd}.

At the angular position $(\theta_0,\varphi_0)$ of the detector, abbreviate the selected-channel normalization $\mathcal{N}_{\nu0}^{({\rm out},\alpha)}$ as $\mathcal{N}_{\nu0}^{(\alpha)}$. The selected global mode then has the near-horizon form
\begin{equation}
u_{\nu}^{\rm out}[x(\tau)]
\simeq\frac{\mathcal{N}_{\nu0}^{(\alpha)}Y_{00}(\theta_0,\varphi_0)}{r_+}
\left[A_{\nu\alpha}^{\rm o}e^{-i\nu u[x(\tau)]}
+A_{\nu\alpha}^{\rm i}e^{-i\nu v[x(\tau)]}\right],
\label{eq:two_branch_mode}
\end{equation}
where $Y_{00}=1/\sqrt{4\pi}$. Equation~\eqref{eq:two_branch_mode} is the leading asymptotic form of one globally normalized mode. A local monopole coupling contains both the logarithmic outgoing phase and the regular ingoing phase.

Frequency selection is implemented by an external transfer function
\begin{equation}
\mathcal{T}_{\rm f}(\nu)=\mathcal{W}_{\rm c}(\nu)
\frac{\Gamma_{\rm c}/2}{\nu-\nu_{\rm c}+i\Gamma_{\rm c}/2},
\qquad
\operatorname{supp}\mathcal{W}_{\rm c}
\subset[\nu_{\min},\nu_{\max}],
\label{eq:frequency_filter}
\end{equation}
with $0<\nu_{\min}<\nu_{\rm c}<\nu_{\max}$. The central frequency $\nu_{\rm c}$ and linewidth $\Gamma_{\rm c}$ are apparatus parameters. This function selects a finite range of Killing frequencies but acts equally on both radial terms in Eq.~\eqref{eq:two_branch_mode}. It does not project onto radial flux.

The band-limited field operator used by the detector is
\begin{equation}
\widehat\Phi_{\rm f}(x)=\int_{\nu_{\min}}^{\nu_{\max}}\dd\nu
\left[\mathcal{T}_{\rm f}(\nu)a_\nu u_\nu^{\rm out}(x)
+\mathcal{T}_{\rm f}^*(\nu)a_\nu^\dagger u_\nu^{\rm out*}(x)\right].
\label{eq:band_limited_field}
\end{equation}
The conjugate transfer function makes $\widehat\Phi_{\rm f}$ Hermitian. This operator represents a selected measurement channel of the scalar field, not a separately normalized field.

We use the finite radial gate
\begin{equation}
\chi_p[x(\tau)]=\Theta(x)
\left(\frac{x}{L_\chi}\right)^p e^{-x/L_\chi},
\qquad
p=0,1,2,\ldots,
\label{eq:gamma_gate}
\end{equation}
where $L_\chi>0$ is a radial length and $\Theta(x)$ is the Heaviside step function that restricts the interaction to the exterior. For $p=0$, the gate is a one-sided exponential. For $p\geq1$, the coupling vanishes at the horizon, and $p\geq2$ also removes the first-derivative jump there. The derivation holds for every nonnegative integer $p$. Define $n=p+1$ and the representative support scale $x_{\rm w}=nL_\chi$. Near-horizon localization requires $x_{\rm w}\ll x_{\rm NH}\ll r_+$, where $x_{\rm NH}$ is a radius within which the expansions in Sec.~\ref{sec:dirac} are accurate. Finite-time detector responses depend on the switching profile, not only on the interaction duration \cite{SriramkumarPadmanabhan:1996,Louko:Satz2008,ShallueCarroll:2025}.

The interaction Hamiltonian is
\begin{equation}
H_{\rm I}(\tau)=g_{\rm c}\,\chi_p[x(\tau)]
\widehat\mu(\tau)\widehat\Phi_{\rm f}[x(\tau)]
\otimes\identity_{\rm spin}.
\label{eq:interaction}
\end{equation}
The coupling $g_{\rm c}$ has the dimensions required by the continuum normalization. Equation~\eqref{eq:interaction} is the standard monopole-detector interaction for a filtered scalar channel \cite{DeWitt:1979QuantumGravity,Scully:2017utk,Louko:Satz2008}. It commutes with the detector's MOG charge and is diagonal in translational spin. At first order, the excitation channel is $|b,0_{\rm B}\rangle\to|a,1_\nu^{\rm out}\rangle$, and the absorption channel is $|b,1_\nu^{\rm out}\rangle\to|a,0_{\rm B}\rangle$.

The phases in Eqs.~\eqref{eq:null_trajectory_expansion} define four positive leading frequency combinations,
\begin{subequations}
\label{eq:phase_frequencies}
\begin{align}
S_{{\rm o},-}&=\frac{\omega}{U_{\rm H}}+\nu C_u,
&S_{{\rm o},+}&=\frac{\omega}{U_{\rm H}}-\nu C_u,
\label{eq:out_phase_frequencies}\\
S_{{\rm i},-}&=\frac{\omega}{U_{\rm H}}+\frac{\nu}{2U_{\rm H}^2},
&S_{{\rm i},+}&=\frac{\omega}{U_{\rm H}}-\frac{\nu}{2U_{\rm H}^2}.
\label{eq:in_phase_frequencies}
\end{align}
\end{subequations}
The subscript $-$ denotes excitation with emission, and $+$ denotes absorption. The label ${\rm o}$ refers to the logarithmic outgoing branch, and ${\rm i}$ refers to the regular ingoing branch. The high-gap analysis requires all four $S$ values to be positive. The convergent integrals remain well defined for either sign because their real damping coefficient is $L_\chi^{-1}$.

The quadratic phase corrections are
\begin{subequations}
\label{eq:quadratic_phase_corrections}
\begin{align}
D_{{\rm o},-}&=\omega T_2+\nu D_u,
\label{eq:quadratic_phase_out_minus}\\
D_{{\rm o},+}&=\omega T_2-\nu D_u,
\label{eq:quadratic_phase_out_plus}\\
D_{{\rm i},-}&=\omega T_2+\nu D_v,
\label{eq:quadratic_phase_in_minus}\\
D_{{\rm i},+}&=\omega T_2-\nu D_v.
\label{eq:quadratic_phase_in_plus}
\end{align}
\end{subequations}
These terms do not enter the leading closed forms, but they determine the phase-truncation error. The ratio $L_\chi/r_+$ alone does not control the approximation. Both branches $b\in\{{\rm o},{\rm i}\}$ must also satisfy $|D_{b,\pm}|\mathfrak m_{b,\pm}^{(2)}\ll1$, where the endpoint moment $\mathfrak m_{b,\pm}^{(2)}$ is defined in Eq.~\eqref{eq:endpoint_moments}.

Define $a_\nu=\nu/\kappa_\alpha$ and
\begin{subequations}
\label{eq:endpoint_parameters}
\begin{align}
\beta_{b,\pm}&=L_\chi^{-1}+iS_{b,\pm},
\label{eq:endpoint_beta}\\
\rho_{b,\pm}&=|\beta_{b,\pm}|,
\label{eq:endpoint_rho}\\
\vartheta_{b,\pm}&=\arg\beta_{b,\pm}=\arctan(S_{b,\pm}L_\chi).
\label{eq:endpoint_angle}
\end{align}
\end{subequations}
At leading order, changing variables from $\tau$ to $x$ gives $\dd\tau=-\dd x/U_{\rm H}$. The four endpoint integrals are
\begin{subequations}
\label{eq:four_integrals}
\begin{align}
J_{{\rm o},-}&=L_\chi^{-p}\Gamma(n-ia_\nu)
\beta_{{\rm o},-}^{-n+ia_\nu},
&J_{{\rm o},+}&=L_\chi^{-p}\Gamma(n+ia_\nu)
\beta_{{\rm o},+}^{-n-ia_\nu},
\label{eq:out_integrals}\\
J_{{\rm i},-}&=L_\chi^{-p}\Gamma(n)\beta_{{\rm i},-}^{-n},
&J_{{\rm i},+}&=L_\chi^{-p}\Gamma(n)\beta_{{\rm i},+}^{-n}.
\label{eq:in_integrals}
\end{align}
\end{subequations}
Here $\Gamma(z)$ is Euler's gamma function. For example, $J_{{\rm o},-}$ follows from $L_\chi^{-p}\int_0^\infty x^{p-ia_\nu}e^{-\beta_{{\rm o},-}x}\dd x$ and the identity $\int_0^\infty x^{z-1}e^{-\beta x}\dd x=\Gamma(z)\beta^{-z}$ for $\operatorname{Re}z>0$ and $\operatorname{Re}\beta>0$. These conditions hold because $n=p+1>0$ and $\operatorname{Re}\beta_{b,\pm}=L_\chi^{-1}>0$. The outgoing term is a Mellin--Laplace transform and contains $\Gamma(n\mp ia_\nu)$, whereas the regular ingoing term is an ordinary Laplace transform.

Oscillatory cancellation requires the approximation error to be measured against the complex endpoint integral, not merely against the positive gate envelope. Define
\begin{equation}
\begin{gathered}
z_{{\rm o},-}=n-ia_\nu,\qquad
z_{{\rm o},+}=n+ia_\nu,\qquad
z_{{\rm i},\pm}=n,\\
\mathfrak{m}_{b,\pm}^{(k)}
=\frac{|(z_{b,\pm})_k|}{\rho_{b,\pm}^{k}},
\qquad
(z)_k\equiv\frac{\Gamma(z+k)}{\Gamma(z)}.
\end{gathered}
\label{eq:endpoint_moments}
\end{equation}
Equation~\eqref{eq:endpoint_moments} gives the magnitude of the $k$th relative moment of the retained linearized transform because inserting $x^k$ multiplies the integral by $(z_{b,\pm})_k/\beta_{b,\pm}^k$. The moment $\mathfrak m^{(1)}$ controls linear amplitude corrections, and $\mathfrak m^{(2)}$ controls the first quadratic-phase correction. These scales retain the dependence on $a_\nu$ that would be lost by replacing each moment with $x_{\rm w}^k$.

A separate tail error arises because the closed integrals extend the near-horizon integrand beyond $x_{\rm NH}$. Let $\mathcal{I}_{b,\pm}^{\rm exact}(x)$ be the exact branch integrand after removing its constant scattering coefficient and the common factor $1/U_{\rm H}$. Let $\mathcal{I}_{b,\pm}^{\rm NH}(x)$ be the corresponding integrand used in Eq.~\eqref{eq:four_integrals}. Define the relative mismatch
\begin{equation}
\epsilon_{\rm tail}
=\max_{b,\pm}
\frac{\left|\int_{x_{\rm NH}}^\infty
[\mathcal{I}_{b,\pm}^{\rm exact}(x)
-\mathcal{I}_{b,\pm}^{\rm NH}(x)]\dd x\right|}
{|J_{b,\pm}|}.
\label{eq:relative_tail_error}
\end{equation}
Equation~\eqref{eq:relative_tail_error} gives the condition required for a noncompact gate. The retained near-horizon part of this tail has the absolute envelope bound $L_\chi\Gamma(n,x_{\rm NH}/L_\chi)$, where $\Gamma(n,z)$ is the upper incomplete gamma function. The ratio of this bound to $|J_{b,\pm}|$ must be small, and the exact exterior integrand must be bounded or checked separately. The positive quantity $Q(n,x_{\rm NH}/L_\chi)=\Gamma(n,x_{\rm NH}/L_\chi)/\Gamma(n)$ alone is not a relative error bound when the denominator is oscillatory.

The common controlled error will be denoted by $\epsilon_{\rm app}$. A sufficient estimate is
\begin{equation}
\begin{aligned}
\epsilon_{\rm app}={}&\order\left[\max_{b,\pm}
\left(\frac{\mathfrak m_{b,\pm}^{(1)}}{r_+}
+\frac{|U_1|\mathfrak m_{b,\pm}^{(1)}}{U_{\rm H}}
+\frac{|f''_+|\mathfrak m_{b,\pm}^{(1)}}{\kappa_\alpha}\right)\right]
+\order\left(\max_{b,\pm}|D_{b,\pm}|\mathfrak m_{b,\pm}^{(2)}\right)\\
&+\order\left(\max_{b,\pm}|c_V^b|\mathfrak m_{b,\pm}^{(1)}\right)
+\order(\epsilon_{\rm tail})
+\order(\epsilon_{\rm rec}).
\end{aligned}
\label{eq:approximation_error}
\end{equation}
The first line of Eq.~\eqref{eq:approximation_error} controls the trajectory, amplitude, and phase expansions through their endpoint moments. The second line controls the local plane-wave form of the scalar branch, the relative exterior tail, and the fixed-worldline approximation. A positive lower frequency $\nu_{\min}$ is required because the scalar wave-zone error is not uniform as $\nu\to0$.

\section{Closed response and detailed balance}
\label{sec:response}

The phases independent of $x$ can be absorbed into two complex scattering coefficients,
\begin{subequations}
\label{eq:scattering_coefficients}
\begin{align}
\mathcal{C}_{\rm o}(\nu,\alpha)
&=\mathcal{N}_{\nu0}^{(\alpha)}A_{\nu\alpha}^{\rm o}
e^{-i\nu u_0}r_+^{-ia_\nu},
\label{eq:outgoing_scattering_coefficient}\\
\mathcal{C}_{\rm i}(\nu,\alpha)
&=\mathcal{N}_{\nu0}^{(\alpha)}A_{\nu\alpha}^{\rm i}e^{-i\nu v_0}.
\label{eq:ingoing_scattering_coefficient}
\end{align}
\end{subequations}
The factor $r_+^{-ia_\nu}$ is a pure phase from $\ln(x/r_+)$ and does not affect a one-branch probability. It does affect the relative phase of the two-branch interference term.

To first order in $g_{\rm c}$, define
\begin{equation}
\mathcal{K}_\nu
=\frac{|g_{\rm c}|^2|\mathcal{T}_{\rm f}(\nu)|^2
|Y_{00}(\theta_0,\varphi_0)|^2}
{\hbar^2U_{\rm H}^2r_+^2}.
\label{eq:response_prefactor}
\end{equation}
The excitation and absorption probability densities into the same asymptotic-out channel are
\begin{subequations}
\label{eq:full_probabilities}
\begin{align}
\frac{\dd P_{\rm exc}}{\dd\nu}
&=\mathcal{K}_\nu
\left|\mathcal{C}_{\rm o}^*J_{{\rm o},-}
+\mathcal{C}_{\rm i}^*J_{{\rm i},-}+\Delta_-\right|^2,
\label{eq:full_excitation}\\
\frac{\dd P_{\rm abs}}{\dd\nu}
&=\mathcal{K}_\nu
\left|\mathcal{C}_{\rm o}J_{{\rm o},+}
+\mathcal{C}_{\rm i}J_{{\rm i},+}+\Delta_+\right|^2.
\label{eq:full_absorption}
\end{align}
\end{subequations}
The remainders satisfy
\begin{equation}
|\Delta_\mp|\lesssim\epsilon_{\rm app}
\left(|\mathcal{C}_{\rm o}J_{{\rm o},\mp}|
+|\mathcal{C}_{\rm i}J_{{\rm i},\mp}|\right).
\label{eq:amplitude_remainder_bound}
\end{equation}
Equations~\eqref{eq:full_excitation} and \eqref{eq:full_absorption} retain the regular ingoing contribution and its coherent interference with the logarithmic outgoing branch. Near an accidental zero from destructive interference, the full probability does not have a useful multiplicative relative error. We therefore state the approximation error at the amplitude level.

Expanding Eq.~\eqref{eq:full_excitation}, for example, gives
\begin{equation}
\begin{aligned}
\frac{1}{\mathcal{K}_\nu}\frac{\dd P_{\rm exc}}{\dd\nu}
={}&|\mathcal{C}_{\rm o}|^2|J_{{\rm o},-}|^2
+|\mathcal{C}_{\rm i}|^2|J_{{\rm i},-}|^2\\
&+2\operatorname{Re}\left(
\mathcal{C}_{\rm o}^*\mathcal{C}_{\rm i}
J_{{\rm o},-}J_{{\rm i},-}^*\right)
+\hbox{controlled remainder}.
\end{aligned}
\label{eq:excitation_decomposition}
\end{equation}
The three terms are the outgoing HBAR contribution, the regular ingoing background, and their interference. The absorption probability has the same structure with $-\to+$ and unconjugated scattering coefficients in the amplitude.

The branch-resolved outgoing densities follow from the gamma-function identity in Eq.~\eqref{eq:out_integrals}. Since
\begin{subequations}
\label{eq:beta_moduli}
\begin{align}
|\beta_{{\rm o},-}^{-n+ia_\nu}|^2
&=\rho_{{\rm o},-}^{-2n}e^{-2a_\nu\vartheta_{{\rm o},-}},
\label{eq:beta_modulus_excitation}\\
|\beta_{{\rm o},+}^{-n-ia_\nu}|^2
&=\rho_{{\rm o},+}^{-2n}e^{2a_\nu\vartheta_{{\rm o},+}}.
\label{eq:beta_modulus_absorption}
\end{align}
\end{subequations}
one obtains
\begin{subequations}
\label{eq:outgoing_densities}
\begin{align}
\frac{\dd P_{\rm exc}^{\rm o}}{\dd\nu}
&=\mathcal{K}_\nu|\mathcal{C}_{\rm o}|^2L_\chi^{-2p}
|\Gamma(n-ia_\nu)|^2
\frac{e^{-2a_\nu\vartheta_{{\rm o},-}}}
{\rho_{{\rm o},-}^{2n}}
\left[1+\order(\epsilon_{\rm app})\right],
\label{eq:out_excitation_density}\\
\frac{\dd P_{\rm abs}^{\rm o}}{\dd\nu}
&=\mathcal{K}_\nu|\mathcal{C}_{\rm o}|^2L_\chi^{-2p}
|\Gamma(n+ia_\nu)|^2
\frac{e^{2a_\nu\vartheta_{{\rm o},+}}}
{\rho_{{\rm o},+}^{2n}}
\left[1+\order(\epsilon_{\rm app})\right].
\label{eq:out_absorption_density}
\end{align}
\end{subequations}
These densities apply when the positive-flux horizon branch is resolved. Their common global normalization, external filter, and coupling cancel at fixed $\nu$.

Because $|\Gamma(n-ia)|=|\Gamma(n+ia)|$, the exact outgoing-branch ratio is
\begin{equation}
\mathcal{R}_{\rm out}(\nu)
\equiv\frac{\dd P_{\rm exc}^{\rm o}/\dd\nu}
{\dd P_{\rm abs}^{\rm o}/\dd\nu}
=\left(\frac{\rho_{{\rm o},+}}{\rho_{{\rm o},-}}\right)^{2n}
\exp\left[-2a_\nu
(\vartheta_{{\rm o},-}+\vartheta_{{\rm o},+})\right]
\left[1+\order(\epsilon_{\rm app})\right].
\label{eq:exact_outgoing_ratio}
\end{equation}
The power $2n=2(p+1)$ in Eq.~\eqref{eq:exact_outgoing_ratio} comes from the radial envelope. The exponential contains the logarithmic horizon phase.

To relate Eq.~\eqref{eq:exact_outgoing_ratio} to the physical two-branch response, define
\begin{subequations}
\label{eq:branch_ratios}
\begin{align}
q_{\rm exc}(\nu)&=
\frac{\mathcal{C}_{\rm i}^*J_{{\rm i},-}}
{\mathcal{C}_{\rm o}^*J_{{\rm o},-}},
\label{eq:q_exc}\\
q_{\rm abs}(\nu)&=
\frac{\mathcal{C}_{\rm i}J_{{\rm i},+}}
{\mathcal{C}_{\rm o}J_{{\rm o},+}}.
\label{eq:q_abs}
\end{align}
\end{subequations}
Away from a zero of an outgoing amplitude, the complete ratio factorizes as
\begin{equation}
\mathcal{R}_{\rm full}(\nu)
\equiv\frac{\dd P_{\rm exc}/\dd\nu}{\dd P_{\rm abs}/\dd\nu}
=\mathcal{R}_{\rm out}(\nu)
\frac{|1+q_{\rm exc}(\nu)|^2}{|1+q_{\rm abs}(\nu)|^2}
+\order(\epsilon_{\rm app}).
\label{eq:full_ratio}
\end{equation}
Equation~\eqref{eq:full_ratio} separates a branch-resolved HBAR measurement from a local coupling to the global mode at leading order. The second factor contains the reflection magnitude, its scattering phase, and the regular ingoing integral. The surface gravity $\kappa_\alpha$ does not determine this factor.

The quality of the one-branch approximation can be bounded without selecting a scattering phase. If $|q_X|\leq\eta_X<1$ for $X\in\{{\rm exc},{\rm abs}\}$, then the difference between the full and outgoing logarithmic balance functions obeys
\begin{equation}
\left|\mathcal{B}_{\rm full}-\mathcal{B}_{\rm out}\right|
\leq-2\ln\left[(1-\eta_{\rm exc})(1-\eta_{\rm abs})\right]
=2(\eta_{\rm exc}+\eta_{\rm abs})+\order(\eta_X^2),
\label{eq:branch_error_bound}
\end{equation}
where $\mathcal{B}=-\ln\mathcal{R}=\ln[(\dd P_{\rm abs}/\dd\nu)/(\dd P_{\rm exc}/\dd\nu)]$. This bound gives a phase-independent condition for the one-branch approximation.

We now isolate the thermal limit of the outgoing branch. For $S_{{\rm o},\pm}L_\chi\gg1$ and $S_{{\rm o},\pm}>0$,
\begin{subequations}
\label{eq:adiabatic_endpoint_expansions}
\begin{align}
\vartheta_{{\rm o},\pm}&=\frac{\pi}{2}
-\frac{1}{S_{{\rm o},\pm}L_\chi}
+\order[(S_{{\rm o},\pm}L_\chi)^{-3}],
\label{eq:adiabatic_angle_expansion}\\
\ln\rho_{{\rm o},\pm}&=\ln S_{{\rm o},\pm}
+\order[(S_{{\rm o},\pm}L_\chi)^{-2}].
\label{eq:adiabatic_rho_expansion}
\end{align}
\end{subequations}
Substitution into Eq.~\eqref{eq:exact_outgoing_ratio} gives the controlled logarithmic expansion
\begin{equation}
\begin{aligned}
\ln\mathcal{R}_{\rm out}={}&-2\pi a_\nu
+2n\ln\left(\frac{S_{{\rm o},+}}{S_{{\rm o},-}}\right)
+\frac{2a_\nu}{L_\chi}
\left(\frac{1}{S_{{\rm o},-}}+\frac{1}{S_{{\rm o},+}}\right)\\
&+\order\left(\sum_{\pm}\frac{n}{(S_{{\rm o},\pm}L_\chi)^2}
+\sum_{\pm}\frac{a_\nu}{(S_{{\rm o},\pm}L_\chi)^3}
+\epsilon_{\rm app}\right).
\end{aligned}
\label{eq:adiabatic_log_ratio}
\end{equation}
The third term in Eq.~\eqref{eq:adiabatic_log_ratio} is proportional to $a_\nu/(S_\pm L_\chi)$. Omitting $a_\nu=\nu/\kappa_\alpha$ underestimates the adiabatic error when the observed mode frequency is not small relative to the surface-gravity scale.

Let $\Omega=\omega/U_{\rm H}$ and define the trajectory or geometric-optics parameter
\begin{equation}
\eta_{\rm go}=\frac{\nu|C_u|}{\Omega}\ll1.
\label{eq:high_gap_parameter}
\end{equation}
Under this condition, $S_{{\rm o},\pm}=\Omega[1+\order(\eta_{\rm go})]$. Combining it with Eq.~\eqref{eq:adiabatic_log_ratio} yields
\begin{equation}
\mathcal{R}_{\rm out}(\nu)
=e^{-2\pi\nu/\kappa_\alpha}
\left[1+\order(n\eta_{\rm go})
+\order\left(\frac{a_\nu}{\Omega L_\chi}\right)
+\order\left(\frac{n}{(\Omega L_\chi)^2}\right)
+\order(\epsilon_{\rm app})\right].
\label{eq:thermal_outgoing_ratio}
\end{equation}
The exponent in Eq.~\eqref{eq:thermal_outgoing_ratio} equals $\hbar\nu/(k_{\rm B}T_{\rm H}^{\rm MOG})$. Thus the direction-resolved outgoing sector obeys Hawking-temperature detailed balance in this limit. The full response has the same limit only if the branch factor in Eq.~\eqref{eq:full_ratio} approaches unity.

For integer $p$, repeated use of $\Gamma(z+1)=z\Gamma(z)$ gives
\begin{subequations}
\label{eq:integer_gate_gamma_identity}
\begin{align}
|\Gamma(n-ia)|^2
&=\Pi_p(a)\frac{\pi a}{\sinh(\pi a)},
\label{eq:gamma_polynomial_identity}\\
\Pi_p(a)&=\prod_{j=1}^{p}(j^2+a^2),
\qquad \Pi_0=1.
\label{eq:gate_polynomial}
\end{align}
\end{subequations}
The leading high-gap outgoing densities are therefore
\begin{subequations}
\label{eq:high_gap_densities}
\begin{align}
\frac{\dd P_{\rm exc}^{\rm o}}{\dd\nu}
&\simeq\frac{|g_{\rm c}|^2|\mathcal{T}_{\rm f}|^2
|\mathcal{C}_{\rm o}|^2|Y_{00}|^2}
{\hbar^2r_+^2\omega^2}
\left(\frac{U_{\rm H}}{\omega L_\chi}\right)^{2p}
\Pi_p(a_\nu)
\frac{2\pi a_\nu}{e^{2\pi a_\nu}-1},
\label{eq:high_gap_excitation}\\
\frac{\dd P_{\rm abs}^{\rm o}}{\dd\nu}
&\simeq\frac{|g_{\rm c}|^2|\mathcal{T}_{\rm f}|^2
|\mathcal{C}_{\rm o}|^2|Y_{00}|^2}
{\hbar^2r_+^2\omega^2}
\left(\frac{U_{\rm H}}{\omega L_\chi}\right)^{2p}
\Pi_p(a_\nu)
\frac{2\pi a_\nu e^{2\pi a_\nu}}{e^{2\pi a_\nu}-1}.
\label{eq:high_gap_absorption}
\end{align}
\end{subequations}
Changing the radial gate modifies the absolute spectral weight through $\Pi_p$ and the factor $(U_{\rm H}/\omega L_\chi)^{2p}$, but it does not change the leading outgoing detailed-balance exponent. For comparison with earlier work, set $p=0$ and $\alpha=0$, define the Schwarzschild radius $r_g=2G_{\rm N}M$, and use $\kappa_0=1/(2r_g)$. The excitation kernel then reduces to $2\pi a_0/(e^{2\pi a_0}-1)=4\pi r_g\nu/(e^{4\pi r_g\nu}-1)$, while $P_{\rm abs}^{\rm o}/P_{\rm exc}^{\rm o}=e^{4\pi r_g\nu}$. After matching the coupling and mode normalization, these expressions agree with the high-gap HBAR probability kernel and detailed-balance relation reported by Scully and collaborators. Camblong and collaborators obtain the same Planck factor from the near-horizon conformal mode \cite{Scully:2017utk,Camblong:2020qtd}. The present calculation adds finite gates with $p>0$, MOG forcing, and coherent branch interference.

Physical probabilities are finite passband integrals,
\begin{equation}
P_X^{\rm f}=\int_{\nu_{\min}}^{\nu_{\max}}\frac{\dd P_X}{\dd\nu}\,\dd\nu,
\qquad X\in\{{\rm exc},{\rm abs}\}.
\label{eq:filtered_probabilities}
\end{equation}
The two probabilities in Eq.~\eqref{eq:filtered_probabilities} refer to different prepared initial states. First-order perturbation theory requires $P_{\rm exc}^{\rm f}\ll1$ for the vacuum-emission experiment and $P_{\rm abs}^{\rm f}\ll1$ for the one-particle absorption experiment. Their sum is not a general perturbative criterion. Finite-time detector probabilities depend on the detailed switching function \cite{SriramkumarPadmanabhan:1996,Louko:Satz2008}.

If the transfer-function-removed density $\dd\overline P_X/\dd\nu\equiv|\mathcal{T}_{\rm f}|^{-2}\dd P_X/\dd\nu$ varies slowly across the filter, the narrow-band approximation applies. Define $\Delta\nu_{\rm eff}=\int_{\nu_{\min}}^{\nu_{\max}}|\mathcal{T}_{\rm f}(\nu)|^2\dd\nu$. Then $P_X^{\rm f}\simeq\Delta\nu_{\rm eff}[\dd\overline P_X/\dd\nu]_{\nu_c}$, provided that
\begin{equation}
\epsilon_{\rm bw}=\Gamma_{\rm c}
\max_{\nu\in[\nu_{\min},\nu_{\max}]}
\left|\partial_\nu\ln(\dd\overline P_X/\dd\nu)\right|\ll1.
\label{eq:narrow_band_condition}
\end{equation}
For a passband extending many linewidths on both sides of a resonance with $\nu_c\gg\Gamma_c$, $\Delta\nu_{\rm eff}\simeq\pi\Gamma_c/2$. The apparatus, not the black hole, sets this bandwidth.

\section{Limits, weak-MOG expansion, and observables}
\label{sec:limits}

Let $s_\alpha=\sqrt{1+\alpha}$. The exact horizon quantities are
\begin{equation}
r_+=G_{\rm N}Ms_\alpha(1+s_\alpha),
\qquad
\frac{\kappa_\alpha}{\kappa_0}
=\frac{4}{s_\alpha(1+s_\alpha)^2},
\qquad
\kappa_0=\frac{1}{4G_{\rm N}M}.
\label{eq:exact_geometry_ratios}
\end{equation}
For $\alpha\geq0$, Eq.~\eqref{eq:exact_geometry_ratios} gives $0<\kappa_\alpha/\kappa_0\leq1$. At fixed asymptotic mass $M$, the Schwarzschild--MOG temperature does not exceed the Schwarzschild value.

The horizon trajectory function and the linear retarded-time coefficient are
\begin{equation}
U_{\rm H}=\varepsilon-\frac{s_\alpha-1}{s_\alpha},
\qquad
C_u=\frac{1}{2U_{\rm H}^2}
-\frac{(s_\alpha-3)(s_\alpha+1)}{2}.
\label{eq:exact_trajectory_coefficients}
\end{equation}
Equation~\eqref{eq:exact_trajectory_coefficients} contains the vector-force correction through $U_{\rm H}$ and the metric correction through the second term. For release from rest at infinity, $\varepsilon=1$, these coefficients reduce to $U_{\rm H}=s_\alpha^{-1}$ and $C_u=s_\alpha+3/2$. Such a packet crosses the horizon for every finite $\alpha\geq0$.

In the Schwarzschild limit, $r_+\to2G_{\rm N}M$, $\kappa_\alpha\to\kappa_0$, $q_\psi\to0$, and $U_{\rm H}\to\varepsilon$. The Dirac system, trajectory, and response then reduce smoothly to their Schwarzschild forms with the same global-mode convention and radial gate. Equation~\eqref{eq:test_charge} ties the charge-to-mass ratio to $\alpha$, so the standard MOG prescription has no separate limit in which $q_\psi$ vanishes while $\alpha$ and the background remain fixed.

The thermal part of the high-gap outgoing excitation density can be compared without specifying the global scattering coefficient or gate-dependent polynomial. Define the normalization-reduced spectral ratio
\begin{equation}
\mathcal{R}_\alpha^{\rm spec}(\nu)
=\frac{\kappa_0}{\kappa_\alpha}
\frac{e^{2\pi\nu/\kappa_0}-1}
{e^{2\pi\nu/\kappa_\alpha}-1}.
\label{eq:reduced_spectral_ratio}
\end{equation}
This ratio isolates the change in the Planck-weighted factor $2\pi a/(e^{2\pi a}-1)$. It is not a ratio of complete detector probabilities.

At fixed $M$ and $\nu$, let $r_{+,0}=2G_{\rm N}M$, $r_{+,\alpha}=r_+$, $U_{{\rm H},0}=\varepsilon$, and
\begin{equation}
\mathcal{G}_\alpha^{\rm o}(\nu)
=\frac{|\mathcal{C}_{\rm o}(\nu,\alpha)|^2}
{|\mathcal{C}_{\rm o}(\nu,0)|^2}
=\frac{|\mathcal{N}_{\nu0}^{(\alpha)}|^2}
{|\mathcal{N}_{\nu0}^{(0)}|^2}
\frac{\Gamma_{\nu0}^{(0)}}{\Gamma_{\nu0}^{(\alpha)}}
\label{eq:global_outgoing_ratio}
\end{equation}
for the unit-outgoing-at-infinity radial convention in Eq.~\eqref{eq:out_mode_boundaries}. The second equality follows from $\Gamma_{\nu0}^{(\alpha)}=|A_{\nu\alpha}^{\rm o}|^{-2}$ and retains the possible background dependence of the Klein--Gordon normalization. It reduces to the inverse-greybody ratio alone only when $|\mathcal{N}_{\nu0}^{(\alpha)}|=|\mathcal{N}_{\nu0}^{(0)}|$ under the chosen continuum convention. The high-gap outgoing excitation ratio for the same $p$, $L_\chi$, $\omega$, coupling, and filter is
\begin{equation}
\mathcal{R}_{\alpha,p}^{\rm o,exc}
=\mathcal{G}_\alpha^{\rm o}
\frac{r_{+,0}^2}{r_{+,\alpha}^2}
\left(\frac{U_{{\rm H},\alpha}}{U_{{\rm H},0}}\right)^{2p}
\frac{\Pi_p(\nu/\kappa_\alpha)}{\Pi_p(\nu/\kappa_0)}
\mathcal{R}_\alpha^{\rm spec}.
\label{eq:outgoing_mog_ratio}
\end{equation}
Equation~\eqref{eq:outgoing_mog_ratio} includes the factors omitted from the reduced ratio. Even for a branch-resolved measurement, the response depends on the horizon radius, MOG-forced trajectory, radial gate, and greybody propagation.

For the local two-branch detector, Eq.~\eqref{eq:full_excitation} adds a further interference factor. The complete MOG-to-Schwarzschild excitation ratio is
\begin{equation}
\mathcal{R}_{\alpha,p}^{\rm full,exc}
=\mathcal{R}_{\alpha,p}^{\rm o,exc}
\frac{|1+q_{{\rm exc},\alpha}|^2}
{|1+q_{{\rm exc},0}|^2}
+\order(\epsilon_{\rm app}).
\label{eq:full_mog_ratio}
\end{equation}
Equation~\eqref{eq:full_mog_ratio} gives the full comparison for the selected asymptotic-out channel. Setting $\mathcal{G}_\alpha^{\rm o}=1$ and $q_{\rm exc}=0$ imposes two independent assumptions about global propagation.

\begin{figure}[t]
\centering
\includegraphics[width=0.82\linewidth]{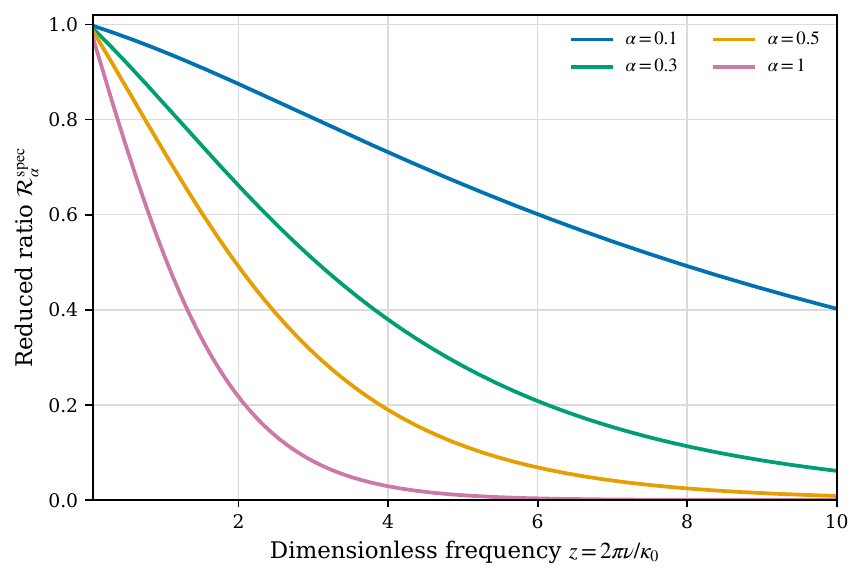}
\caption{Normalization-reduced thermal spectral ratio in Eq.~\eqref{eq:reduced_spectral_ratio} as a function of $z=2\pi\nu/\kappa_0$. The ratio omits the gate polynomial, horizon-radius normalization, global scattering coefficient, and two-branch interference.}
\label{fig:reduced_spectral_ratio}
\end{figure}

Figure~\ref{fig:reduced_spectral_ratio} shows the reduced quantity in Eq.~\eqref{eq:reduced_spectral_ratio}. Its monotonic decrease follows from the decrease of $\kappa_\alpha$ with $\alpha$. The full probability also requires the factors in Eqs.~\eqref{eq:outgoing_mog_ratio} and \eqref{eq:full_mog_ratio}.

For $\alpha\ll1$, the geometry and trajectory coefficients are
\begin{subequations}
\label{eq:weak_geometry_trajectory_expansions}
\begin{align}
\frac{r_+}{2G_{\rm N}M}&=1+\frac34\alpha-\frac1{16}\alpha^2+\order(\alpha^3),
\label{eq:weak_horizon_radius}\\
\frac{\kappa_\alpha}{\kappa_0}&=1-\alpha+\frac{15}{16}\alpha^2+\order(\alpha^3),
\label{eq:weak_surface_gravity}\\
U_{\rm H}&=\varepsilon-\frac12\alpha+\frac38\alpha^2+\order(\alpha^3),
\label{eq:weak_horizon_trajectory}\\
C_u&=2+\frac{1}{2\varepsilon^2}
+\frac{\alpha}{2\varepsilon^3}+\order(\alpha^2).
\label{eq:weak_retarded_coefficient}
\end{align}
\end{subequations}
These expansions keep $M$, $\varepsilon$, $\omega$, $\nu$, and $L_\chi$ fixed. Writing $z=2\pi\nu/\kappa_0$, Eq.~\eqref{eq:reduced_spectral_ratio} becomes
\begin{equation}
\mathcal{R}_\alpha^{\rm spec}(z)
=1+\alpha\left[1-\frac{z}{1-e^{-z}}\right]
+\order(\alpha^2).
\label{eq:weak_spectral_ratio}
\end{equation}
The coefficient in Eq.~\eqref{eq:weak_spectral_ratio} is negative for every $z>0$. A positive weak MOG deformation therefore suppresses the reduced thermal spectrum at fixed $M$ and $\nu$.

The finite-gate balance contains more information. Define the outgoing logarithmic balance
\begin{equation}
\mathcal{B}_{\rm out}(\nu,\alpha)
=2n\ln\left(\frac{\rho_{{\rm o},-}}{\rho_{{\rm o},+}}\right)
+2a_\nu(\vartheta_{{\rm o},-}+\vartheta_{{\rm o},+}).
\label{eq:outgoing_balance}
\end{equation}
At leading near-horizon order, Eq.~\eqref{eq:outgoing_balance} equals $-\ln\mathcal{R}_{\rm out}$. It separates the envelope prefactor from the two switching angles.

Write $a_\nu=a_0+\alpha a_1+\cdots$ and $S_{{\rm o},\pm}=S_\pm^{(0)}+\alpha S_\pm^{(1)}+\cdots$. The coefficients needed below are
\begin{subequations}
\label{eq:weak_expansion_coefficients}
\begin{align}
a_0&=\frac{\nu}{\kappa_0},
\label{eq:weak_a0}\\
a_1&=a_0,
\label{eq:weak_a1}\\
S_-^{(0)}&=\frac{\omega}{\varepsilon}
+\nu\left(2+\frac{1}{2\varepsilon^2}\right),
\label{eq:weak_S_minus_zero}\\
S_+^{(0)}&=\frac{\omega}{\varepsilon}
-\nu\left(2+\frac{1}{2\varepsilon^2}\right),
\label{eq:weak_S_plus_zero}\\
S_-^{(1)}&=\frac{\omega}{2\varepsilon^2}
+\frac{\nu}{2\varepsilon^3},
\label{eq:weak_S_minus_one}\\
S_+^{(1)}&=\frac{\omega}{2\varepsilon^2}
-\frac{\nu}{2\varepsilon^3}.
\label{eq:weak_S_plus_one}
\end{align}
\end{subequations}
Let $\rho_\pm^{(0)}=[(S_\pm^{(0)})^2+L_\chi^{-2}]^{1/2}$ and $\vartheta_\pm^{(0)}=\arctan(S_\pm^{(0)}L_\chi)$. Differentiating Eq.~\eqref{eq:outgoing_balance} gives
\begin{equation}
\mathcal{B}_{\rm out}^{(1)}
=\mathcal{B}_{\kappa}^{(1)}
+\mathcal{B}_{\rm traj}^{(1)}
+\mathcal{B}_{\rm gate}^{(1)},
\label{eq:local_weak_decomposition}
\end{equation}
with
\begin{subequations}
\label{eq:local_weak_terms}
\begin{align}
\mathcal{B}_{\kappa}^{(1)}
&=2a_0(\vartheta_-^{(0)}+\vartheta_+^{(0)}),
\label{eq:weak_surface_gravity_term}\\
\mathcal{B}_{\rm traj}^{(1)}
&=2n\left[
\frac{S_-^{(0)}S_-^{(1)}}{(\rho_-^{(0)})^2}
-\frac{S_+^{(0)}S_+^{(1)}}{(\rho_+^{(0)})^2}
\right],
\label{eq:weak_trajectory_term}\\
\mathcal{B}_{\rm gate}^{(1)}
&=\frac{2a_0}{L_\chi}\left[
\frac{S_-^{(1)}}{(\rho_-^{(0)})^2}
+\frac{S_+^{(1)}}{(\rho_+^{(0)})^2}
\right].
\label{eq:weak_gate_term}
\end{align}
\end{subequations}
The first term in Eq.~\eqref{eq:local_weak_decomposition} comes from $a_\nu=\nu/\kappa_\alpha$. The second is the MOG-trajectory correction to the radial prefactor. The third carries the same trajectory deformation through the finite-gate angles. This decomposition depends on the protocol but is fixed once $\chi_p$ and the quantities held constant in the $\alpha$ expansion are specified.

The full logarithmic balance is
\begin{equation}
\mathcal{B}_{\rm full}=\mathcal{B}_{\rm out}
+2\ln|1+q_{\rm abs}|-2\ln|1+q_{\rm exc}|.
\label{eq:full_logarithmic_balance}
\end{equation}
If $q_X=q_X^{(0)}+\alpha q_X^{(1)}+\cdots$, its first-order coefficient contains the additional global term
\begin{equation}
\mathcal{B}_{\rm full}^{(1)}
=\mathcal{B}_{\kappa}^{(1)}
+\mathcal{B}_{\rm traj}^{(1)}
+\mathcal{B}_{\rm gate}^{(1)}
+2\operatorname{Re}\left[
\frac{q_{\rm abs}^{(1)}}{1+q_{\rm abs}^{(0)}}
-\frac{q_{\rm exc}^{(1)}}{1+q_{\rm exc}^{(0)}}
\right].
\label{eq:full_weak_decomposition}
\end{equation}
The last term in Eq.~\eqref{eq:full_weak_decomposition} contains the $\alpha$ dependence of the reflection magnitude and phase and of the regular ingoing integral. Evaluating it requires the complete scalar scattering solution. Replacing only the surface gravity retains the first term and omits the other three.

We leave the three local coefficients in analytic form. In the controlled high-gap and strongly localized regime, they vary only weakly over a narrow passband. The global term cannot be evaluated without solving the scattering problem.

The low-frequency limit $\mathcal{R}_\alpha^{\rm spec}\to1$ applies only to the reduced thermal factor. The near-horizon plane-wave approximation is not uniform in this limit unless $\nu$ remains above the positive lower edge of the filter. At high frequency,
\begin{equation}
\mathcal{R}_\alpha^{\rm spec}\sim\frac{\kappa_0}{\kappa_\alpha}
\exp\left[-2\pi\nu\left(\kappa_\alpha^{-1}-\kappa_0^{-1}\right)\right],
\label{eq:high_frequency_spectral_ratio}
\end{equation}
so the reduced MOG spectrum has stronger exponential suppression for $\alpha>0$. Formally, $r_+\sim G_{\rm N}M\alpha$ and $\kappa_\alpha\sim(G_{\rm N}M\alpha^{3/2})^{-1}$ as $\alpha\to\infty$. This behavior is not a finite-parameter extremal limit. Statements about the full response at large $\alpha$ require scattering data and a check of the fixed-background approximation.

Several observables can be formed without the overall coupling. The local fitted inverse temperature is obtained from the slope
\begin{equation}
m_{\mathcal{B}}(\nu)=\partial_\nu\mathcal{B}(\nu),
\qquad
\kappa_{\rm fit}(\nu)=\frac{2\pi}{m_{\mathcal{B}}(\nu)},
\qquad
T_{\rm fit}(\nu)=\frac{\hbar\kappa_{\rm fit}(\nu)}{2\pi k_{\rm B}}.
\label{eq:fitted_temperature}
\end{equation}
The fitted quantities define a positive temperature when $m_{\mathcal B}>0$. For the outgoing branch, Eq.~\eqref{eq:fitted_temperature} approaches $\kappa_\alpha$ and $T_{\rm H}^{\rm MOG}$ under the conditions of Eq.~\eqref{eq:thermal_outgoing_ratio}. For the full response, the frequency dependence also probes $q_{\rm exc}$ and $q_{\rm abs}$. A measured thermal slope is consistent with the nondegenerate-horizon mechanism but is not a unique signature of MOG.

The finite-band logarithmic balance $\mathcal{B}_{\rm f}=\ln(P_{\rm abs}^{\rm f}/P_{\rm exc}^{\rm f})$ is more directly measurable than a monochromatic density. In the narrow-band limit, it approaches $\mathcal{B}(\nu_c)$ up to $\order(\epsilon_{\rm bw})$. Its interpretation requires a specified field state, passband, gate, detector gap, and global scattering channel.

Table~\ref{tab:validity} lists the independent assumptions. The near-horizon metric expansion alone is insufficient for the Planckian result.

\begin{table}[t]
\caption{Control conditions for the derived responses. Here $x_{\rm w}=(p+1)L_\chi$ and $b\in\{{\rm o},{\rm i}\}$.}
\label{tab:validity}
\begin{ruledtabular}
\begin{tabular}{p{0.27\linewidth}p{0.61\linewidth}}
Approximation & Required control \\
\midrule
Fixed exterior background & Detector and emitted energies are negligible relative to $M$, with no backreaction on $g_{\mu\nu}$ or $\phi_\mu$ \\
Physical Dirac infall & $U_{\rm H}>0$, $E_{\rm H}=m_\psi U_{\rm H}>0$ \\
Dirac horizon reduction & Eq.~\eqref{eq:dirac_control} at $x\lesssim x_{\rm w}$ \\
Localized radial gate & $x_{\rm w}\ll x_{\rm NH}\ll r_+$ and $\epsilon_{\rm tail}\ll1$ from Eq.~\eqref{eq:relative_tail_error} \\
Scalar horizon wave zone & $\max_{b,\pm}|c_V^b|\mathfrak m_{b,\pm}^{(1)}\ll1$ from Eq.~\eqref{eq:scalar_horizon_correction} \\
Endpoint expansion & The first line of Eq.~\eqref{eq:approximation_error} is $\ll1$ \\
Linear phase & $\max_{b,\pm}|D_{b,\pm}|\mathfrak m_{b,\pm}^{(2)}\ll1$ \\
Fixed detector worldline & $\epsilon_{\rm rec}=\hbar\omega/m_\psi\ll1$ \\
Adiabatic outgoing response & $S_{{\rm o},\pm}L_\chi\gg1$ and $a_\nu/(S_{{\rm o},\pm}L_\chi)\ll1$ \\
High-gap response & $\eta_{\rm go}=\nu|C_u|U_{\rm H}/\omega\ll1$ \\
One-branch thermality & $|q_{\rm exc}|\ll1$ and $|q_{\rm abs}|\ll1$, or an independently specified direction-resolved measurement \\
Narrow filter & $0<\nu_{\min}<\nu_c<\nu_{\max}$ and $\epsilon_{\rm bw}\ll1$ \\
First-order interaction & $P_X^{\rm f}\ll1$ for each prepared initial state $X$ \\
\end{tabular}
\end{ruledtabular}
\end{table}

\section{Discussion and conclusion}
\label{sec:discussion}

The leading outgoing result agrees with the HBAR mechanism. An outgoing Killing-frequency mode acquires a logarithmic phase because the retarded coordinate diverges as $-\kappa_\alpha^{-1}\ln(r-r_+)$. In the adiabatic high-gap limit, the Mellin part of the detector integral produces the factor $e^{-2\pi\nu/\kappa_\alpha}$ \cite{Scully:2017utk,Camblong:2020qtd,Sen:2022static}. For a nondegenerate static horizon, this exponent depends on the local geometry through the surface gravity. In the present background, MOG enters this term through $\kappa_\alpha$.

The full detector response also depends on global propagation. A mode that is outgoing at infinity has two asymptotic branches near the horizon. Equation~\eqref{eq:full_probabilities} retains both, and Eq.~\eqref{eq:full_ratio} gives the change in detailed balance from the regular branch. This change is not a greybody multiplier applied after calculating a local probability. It is interference at the amplitude level and depends on the phase of the global scattering solution. Treating a local outgoing ansatz as a complete mode hides this distinction \cite{Candelas:1980,Law:2022Scattering}.

A physical cavity or direction-sensitive detector can suppress one branch, but the suppression must follow from its interaction or boundary conditions. A Lorentzian transfer function alone cannot suppress a radial branch because it acts on frequency, not radial flux. Quasinormal-mode resonances are different. Their centers and widths follow from complex black-hole frequencies rather than external filter parameters \cite{Ovgun:2026QNMHBAR}. Here the apparatus sets the filter, and we do not infer a quasinormal frequency from it.

Finite switching also changes the response. The gate in Eq.~\eqref{eq:gamma_gate} gives closed expressions for each $p$ and removes the abrupt horizon endpoint when $p\geq1$. At finite $L_\chi$, the outgoing ratio in Eq.~\eqref{eq:exact_outgoing_ratio} is nonthermal. Equation~\eqref{eq:adiabatic_log_ratio} shows that the leading angular correction contains the factor $a_\nu=\nu/\kappa_\alpha$. It can remain relevant even when $S_\pm L_\chi$ is large. This sensitivity to the switching function also appears in finite detector measurements near horizons \cite{Louko:Satz2008,ShallueCarroll:2025}.

The MOG vector field changes both the temperature and the detector trajectory. Its matter charge produces the trajectory function $U(r)$ in Eq.~\eqref{eq:trajectory_U}. The identity $E_{\rm H}=m_\psi U_{\rm H}$ connects the Dirac propagation to the WKB trajectory without equating the spinor energy with the internal detector gap. The formal threshold $E_{\rm H}=0$ defines a regular stationary boundary problem for the coupled radial spinor. It lies outside the future-directed HBAR sector because an exterior timelike trajectory requires $U_{\rm H}>0$.

In the weak-MOG expansion, Eq.~\eqref{eq:local_weak_decomposition} contains the surface-gravity deformation, the force-induced change in the radial prefactor, and the finite-gate angle correction. Equation~\eqref{eq:full_weak_decomposition} adds the derivative of the global two-branch factor. Replacing $\kappa_0$ with $\kappa_\alpha$ in a Schwarzschild thermal law gives only the first term. The other terms require the MOG trajectory and measurement protocol. The last term also requires the complete scalar scattering solution.

These results do not give an astrophysical event rate. Such a rate would require a detector injection model, state preparation, coupling calibration, collection efficiency, and scalar scattering data. The reduced ratio in Eq.~\eqref{eq:reduced_spectral_ratio} compares the thermal factors after removing terms retained by a measurement. Equations~\eqref{eq:outgoing_mog_ratio} and \eqref{eq:full_mog_ratio} display those terms without assigning unspecified values to them.

The detector model has a limited microscopic scope. The Dirac equation describes translational propagation, while the two-level monopole is an independent effective degree of freedom. The coupling is neutral and spin preserving. Coupling a detector to a quantized Dirac field would require a Grassmann-even operator and fermionic correlation functions \cite{Louko:2016FermionDetector}. A charged emitted field would require charge conservation across the internal and radiative sectors. A recoil-sensitive composite detector would require a state-dependent stress-energy tensor and MOG current. Changing a frequency in the present scalar formulas does not describe any of these cases.

Under the conditions in Table~\ref{tab:validity}, the MOG-charged Dirac packet has the inverse-square horizon form in Eq.~\eqref{eq:dirac_cqm}, with a positive physical index fixed by Eq.~\eqref{eq:horizon_identity}. Equation~\eqref{eq:full_probabilities} gives the finite-gate excitation and absorption densities for both branches. Their factorization in Eq.~\eqref{eq:full_ratio} separates the outgoing HBAR factor from interference with the regular branch. The weak-MOG balance in Eq.~\eqref{eq:full_weak_decomposition} further separates local geometric and trajectory terms from global propagation. Near-horizon analysis determines the local terms, while a full measurement model and scalar scattering solution must supply the remaining terms.

Further work should solve Eq.~\eqref{eq:scalar_radial} for the complex coefficients $A_{\nu\alpha}^{\rm o}$ and $A_{\nu\alpha}^{\rm i}$, use wave-packet rather than monochromatic final states, and evaluate the response for a microscopic direction-sensitive coupling. These steps would provide the scattering and detector data needed for a full asymptotic prediction.

\begin{acknowledgments}
N.J.L. Lobos and E.T. Rodulfo acknowledge institutional support from De La Salle University and the DLSU Theoretical Physics Group, and research support from the Department of Science and Technology--Accelerated Science and Technology Human Resource Development Program (DOST--ASTHRDP).
\end{acknowledgments}

\section*{Data availability}
No datasets were generated or analyzed in this theoretical study. The equation used to generate the figure is given in the text.

\section*{Competing interests}
The authors declare no competing interests.

\bibliography{ref}

\end{document}